\documentclass[twocolumn,prb,showpacs]{revtex4}
\usepackage{graphicx}
\usepackage{dcolumn}
\usepackage{bm}
\begin{document}
\title{Charge and spin excitation spectra in the one-dimensional 
Hubbard model with next-nearest-neighbor hopping}
\author{S.~Nishimoto,$^{1}$ T.~Shirakawa,$^{2}$ and Y.~Ohta$^{2,3}$}
\affiliation{$^{1}$Max-Planck-Institut f\"ur Physik komplexer Systeme, 
D-01187 Dresden, Germany\\
$^{2}$Graduate School of Science and Technology, Chiba University, 
Chiba 263-8522, Japan\\
$^{3}$Department of Physics, Chiba University, Chiba 263-8522, Japan}
\date{\today}
\begin{abstract}
The dynamical density-matrix renormalization group technique 
is used to calculate spin and charge excitation spectra in the 
one-dimensional (1D) Hubbard model at quarter filling with 
nearest-neighbor $t$ and next-nearest-neighbor $t'$ hopping 
integrals.  We consider a case where $t$ ($>0$) is much smaller 
than $t'$ ($>0$).  
We find that the spin and charge excitation spectra come from 
the two nearly independent $t'$-chains and are basically the 
same as those of the 1D Hubbard (and $t$-$J$) chain at quarter 
filling.  
However, we find that the hopping integral $t$ plays a 
crucial role in the short-range spin and charge correlations; 
i.e., the ferromagnetic spin correlations between electrons 
on the neighboring sites is enhanced and simultaneously the 
spin-triplet pairing correlations is induced, of which the 
consequences are clearly seen in the calculated spin and charge 
excitation spectra at low energies.  
\end{abstract}
\maketitle

\section{INTRODUCTION}

Spin-triplet superconductivity has been one of the major 
issues in the field of condensed-matter physics.  Nearly 
all the conventional and unconventional superconductors known 
to date are spin-singlet paired. The best-known example 
of triplet pairing is not a superconductor but a superfluid 
$^3$He, where atomic Cooper pairs are formed in spin-triplet 
channel.\cite{leggett}  Only a few materials of spin-triplet 
superconductivity have so far been confirmed in strongly 
correlated electron systems, which iclude ruthenium oxide 
Sr$_2$RuO$_4$\cite{maeno} and some heavy-fermion compounds 
such as UPt$_3$.\cite{joynt}  Here, some questions will 
naturally arise.  One is whether the electron correlation 
can take an essential part in superconductivity carried by 
spin-triplet pairs. Another is how the behavior differs from 
that of spin-singlet superconductivity.  In this manner, 
research on spin-triplet superconductivity may offer an 
opportunity to expose unknown physical phenomena.  

Quite recently, a new mechanism of the spin-triplet 
superconductivity has been proposed in a fairly simple 
correlated electron system.\cite{ohta}  The model consistes 
of two Hubbard chains coupled with zigzag bonds and has a unique 
structure of hopping integrals: sign of the hopping integrals 
changes alternately along the zigzag bonds connecting two 
chains, while the sign along the one-dimensional (1D) chain 
is always negative.  [A model where all the hopping integrals 
are taken to be positive is equivalent under canonical 
transformation.]  
Under this sign rule of the hopping integrals, the 
ring-exchange mechanism\cite{fazekas} yields ferromagnetic 
spin correlations, and accordingly, attractive interaction 
between electrons is derived.  In Ref.~\onlinecite{ohta}, 
the argument was developed on the basis of only the static 
properties such as pair binding energy, spin excitation gap, 
and pair correlation function, as well as spin correlation 
function.  Therefore, further investigations including 
dynamical properties have been desired.  

As for real materials, this mechanism\cite{ohta} may be 
of possible relevance to superconductivity observed in 
quasi-1D organic conductor (TMTSF)$_2$X 
[X=PF$_6$, ClO$_4$, etc.], the so-called 
Bechgaard salts.\cite{jerome,bechgaard}  The system exhibits 
a rich phase diagram upon variation of the pressure and 
temperature.  At low temperatures, the phase changes, 
in the order, as the spin-Peierls insulator, 
antiferromagnetic insulator, spin-density-wave (SDW) 
insulator, superconductivity, and paramagnetic metal, 
with increasing pressure.  So far, experimental evidences 
that the superconducting state occurs in the triplet channel 
have been piled up,\cite{lee} although not much is known 
on the nature of the pairing.  
A newly synthesized copper-oxide compound 
Pr$_2$Ba$_4$Cu$_7$O$_{15-\delta}$\cite{matsukawa} 
may also be a relating system.  This material consists 
of both the single CuO chains (as in PrBa$_2$Cu$_3$O$_7$) 
and the double CuO chains (as in PrBa$_2$Cu$_4$O$_8$), and 
those chains are separated by insulating CuO$_2$ plains.  
It has been reported that the double chains turn into a 
superconducting state below $T_{\rm c} \sim 10$ K.\cite{sasaki}  
Although the signs of the hopping integrals seem not to 
satisfy the ferromagnetic sign rule, the structure of 
the double CuO chains bears a certain similarity to our 
model.\cite{sano}  

The purpose of the present study is therefore to build up 
understanding of the ring-exchange superconducting mechanism 
by calculating dynamical quantities for the same model proposed 
in Ref.~\onlinecite{ohta}.  To see the excitations in the spin 
and charge degrees of freedom separately, we here calculate 
the momentum-dependent dynamical spin-spin and density-density 
correlation functions.  We use the dynamical density-matrix 
renormalization group (DDMRG) method for calculating dynamical 
quantities,\cite{eric} which is an extension of the standard 
DMRG method.\cite{white}  The obtained results with high 
resolutions enable us to discuss details of the fundamental 
properties on the spin and charge excitations.  Thus, we can 
find some interesting features in the low-energy physics of 
our model.  

We will show that, although the spin and charge excitation 
spectra are basically the same as those of the two weakly-coupled 
1D Hubbard (and $t$$-$$J$) chains at quarter filling, the small 
hopping integral between the chains plays a crucial role in 
the short-range correlations and low-energy excitaions; i.e., 
the ferromagnetic spin correlations between electrons on the 
neighboring sites is enhanced and the spin-triplet pairing 
correlations between the electrons is induced, of which the 
consequences are clearly seen in the calculated spin and charge 
excitation spectra at low energies.  
We hope that the present investigation will provide deeper 
insight into the mechanism of the spin-triplet superconductivity.

Our paper is organized as follows.  In Sec.~II, we define the 
1D $t$-$t'$-$U$ model and introduce the physical quantities 
of interest, namely, spin and charge excitation spectra.  
In Sec.~III, exact solution of the spin (and charge) excitation 
spectrum in the noninteracting case is presented, and by 
comparing our result with the exact one, we evaluate the 
performance of our DDMRG method.  We then study the spectra 
with onsite Coulomb interaction and discuss relevance to the 
spin-triplet superconductivity.  
We close with a summary in Sec.~IV.

\section{MODEL AND METHOD}

\begin{figure}[tbh]
\begin{center}
\includegraphics[width= 5.0cm]{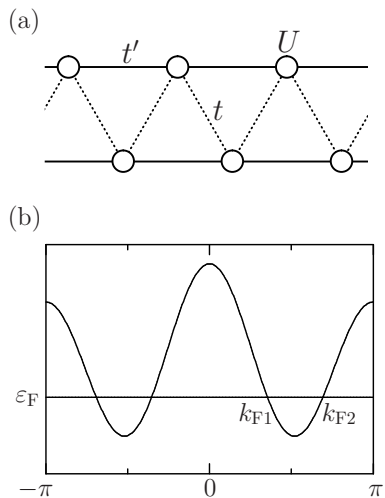}
\end{center}
\caption{Schematic representations of (a) the lattice structure 
of our model and (b) noninteracting band dispersion at 
$t'/t>1$.}
\label{fig0}
\end{figure}

We consider the 1D Hubbard model defined by the Hamiltonian 
\begin{eqnarray}
\nonumber
H=&&t\sum_{i,\sigma}(c^\dagger_{i+1\sigma}c_{i\sigma}+{\rm H.c.})\\
&+&t'\sum_{i,\sigma}(c^\dagger_{i+2\sigma}c_{i\sigma}+{\rm H.c.}) 
+U\sum_in_{i\uparrow}n_{i\downarrow}.
\label{hamiltonian}
\end{eqnarray}
where $c_{i \sigma}^\dagger$ ($c_{i \sigma}$) is the creation 
(anihilation) operator of an electron with spin $\sigma$ 
$(\sigma=\uparrow,\downarrow)$ at site $i$, 
$n_{i\sigma}=c_{i\sigma}^\dagger c_{i\sigma}$ 
is the number operator, $t'$ and $t$ are the nearest-neighbor 
and next-nearest-neighbor hopping integrals respectively, 
and $U$ is the on-site Coulomb interaction.  
We choose the signs of the hopping integrals $t$ and $t'$ 
to be positive, so that the two spins on a triangle 
of the lattice align ferromagnetically when $U$ is 
large.\cite{ohta}  
The lattice structure of our model is shown schematically 
in Fig.~\ref{fig0}(a).  
We call the chain along the $t$ ($t'$) hopping integral 
the $t$-chain ($t'$-chain).  
The dispersion relation is given by
\begin{equation}
\varepsilon_k = 2t \cos ka + 2t' \cos 2ka,
\label{dispersion}
\end{equation}
where $a$ is the lattice constant along the $t$-chain 
(we set $a=1$ hereafter). 
We consider the case where there are four Fermi momenta 
$\pm k_{F1}$ and $\pm k_{F2}$ ($|k_{F2}|>|k_{F1}|$); i.e., 
the case where there are two branches in the noninteracting 
band dispersion [see Fig.~\ref{fig0}(b)].  
This case occurs when the hopping integrals satisfy 
the condition 
\begin{equation} 
\frac{t'}{t}>\frac{\cos^2\big[(2-\rho)\pi/2\big]}
{\sin^2\big[(2-\rho)\pi\big]}
\end{equation}
where $\rho$ is the band filling.  
In this paper, we restrict ourselves to the case where 
$t'$ is a few times as large as $t$ and the system 
is at quarter filling, $\rho=1/2$.  
Hence, the model can be regarded as a double $t'$-chain 
Hubbard model weakly coupled by the $t$-chain.  

Because $t$ is much smaller than $t'$, it would be 
very useful to allow a case of $t = 0$ for familiarization 
with our results.  In the limit of $t \to 0$, the system is 
equivalent to the two independent 1D Hubbard chains at quarter 
filling since electrons are distributed equally to the two 
chains.  The noninteracting band dispersion reads 
$\varepsilon_k=2t'\cos 2k$ and there are four Fermi momenta 
$\pm k_{\rm F1}=\pm 5\pi/8$ and $\pm k_{\rm F2}=\pm 3\pi/8$ 
in the original Brillouin zone defined for the $t$-chain.  
We define $2k_{\rm F}^*=k_{\rm F2}-k_{\rm F1}=\pi/4$ 
as the `nesting' vector in our model at $t=0$.  
We use this definition of the Brillouin zone throughout 
the paper.  

We calculate the spin excitation spectrum
\begin{equation}
S(q,\omega)=\frac{1}{\pi}{\rm Im}
\langle \Psi_0|
s^+_q \frac{1}{\hat{H}+\omega-E_0-{\rm i}\eta}
s^-_{-q}
|\Psi_0 \rangle\, ,
\label{spinspectrum}
\end{equation}
defined with 
$s^+_q=(1/\sqrt{L})\sum_re^{iqr}c^\dagger_{r\uparrow}c_{r\downarrow}$, 
and the charge excitation spectrum
\begin{equation}
N(q,\omega)=\frac{1}{\pi}{\rm Im}
\langle \Psi_0|
n_q \frac{1}{\hat{H}+\omega-E_0-{\rm i}\eta}
n_{-q}
|\Psi_0 \rangle\, ,
\label{chargespectrum}
\end{equation}
defined with $n_q=(1/\sqrt{L})\sum_{r\sigma}e^{iqr}n_{r\sigma}$.  
Here, $|\Psi_0\rangle$ and $E_0$ are, respectively, the wavefunction 
and energy of the ground state of the Hamiltonian 
Eq.~(\ref{hamiltonian}).  
The DDMRG technique is applied to calculate the excitation spectra.  
We here use the open-end boundary condition (OBC) for accurate 
calculation, because the system is relatively hard to deal with 
by the DMRG method due to large long-range hopping integrals.\cite{white} 
When the OBC is used, we need to use the quasimomenta $k=\pi m/(L+1)$ 
for integers $1\leq m\leq L$ on a chain with $L$ sites in order 
to express the momentum-dependent operators $s^+_q$ and 
$n_{q\sigma}$.\cite{holger}

\begin{figure}[t]
\begin{center}
\includegraphics[height=12.0cm]{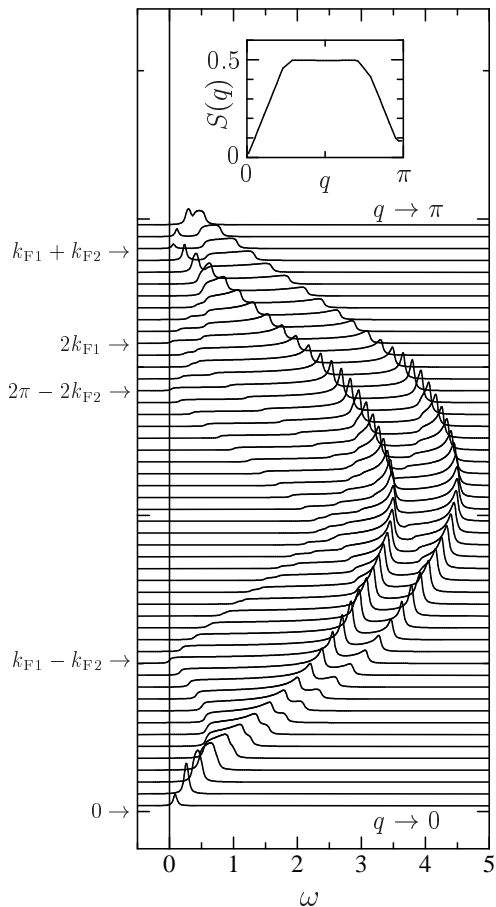}
\end{center}
\caption{Exact spin excitation spectrum $2S(q,\omega)$ [$=N(q,\omega)$] 
for $t=0.25$ and $t'=1$ in the noninteracting case ($U=0$).  
Broadening of the spectrum $\eta=0.03$ is introduced.  
Five momenta indicated with arrows in the left side denote gapless 
points.  Inset: Exact spin structure factor $S(q)$.}
\label{fig1}
\end{figure}

\section{RESULTS}

\subsection*{Noninteracting spectrum}

First, let us consider the noninteracting case, $U=0$, where 
the model is exactly solvable.  In this case, an excitation 
corresponds to creating a particle-hole pair for the ground 
state and therefore we can obtain the exact spectrum of spin 
excitations: 
\begin{equation}
S(q,\omega)=\lim_{\eta \to +0} \frac{1}{\pi L} 
\sum_{\varepsilon_k<\varepsilon_{\rm F}<\varepsilon_{k+q}} 
\frac{\eta}{(\omega-\varepsilon_{k+q}+\varepsilon_k)^2+\eta^2}
\label{exactspec}
\end{equation}
where small $\eta$ is introduced to regularize the poles at 
particular frequencies $\omega$.  Note that the spectrum of 
charge excitations at $U=0$ is exactly twice as large as 
that of the spin excitations, i.e., $N(q,\omega)=2S(q,\omega)$, 
because $N(q,\omega)$ is just a sum over both up and down spins.

\begin{figure}[t]
\begin{center}
\includegraphics[height=12.0cm]{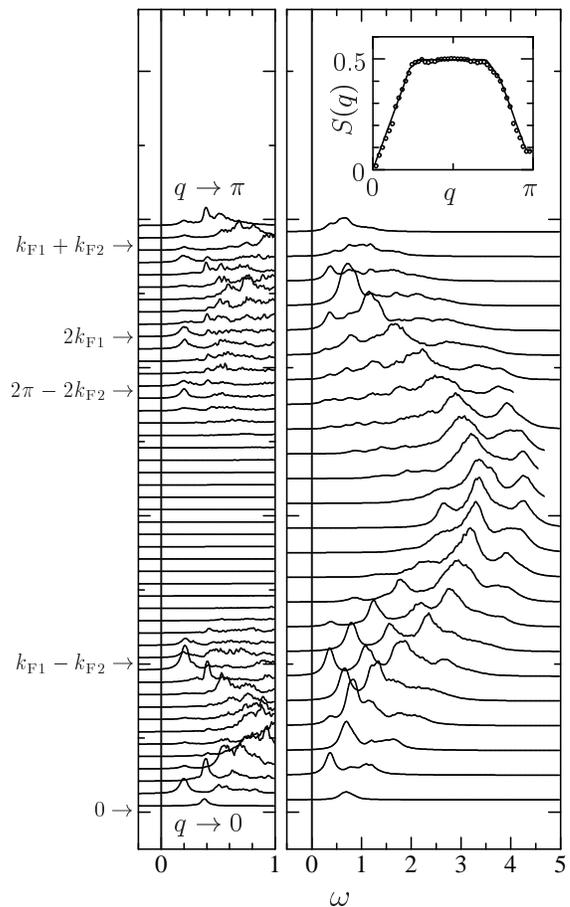}
\end{center}
\caption{Spin excitation spectrum $2S(q,\omega)$ [$=N(q,\omega)$] 
in the noninteracting case ($U=0$) calculated by the DDMRG method.  
The same set of the parameter values as in Fig.~\ref{fig1} is used.  
Right (left) panel shows the result for $L=24$ and $\eta=0.1$ 
($L=48$ and $\eta=0.04$).  
Inset: Spin structure factor $S(q)$ obtained from $\omega$-integration 
of $S(q,\omega)$.}
\label{fig2}
\end{figure}

In Fig.~\ref{fig1}, we show the exact noninteracting spin excitation 
spectrum $S(q,\omega)$ given in Eq.~(\ref{exactspec}).  For small 
$t/t'$ $(<1)$, the spectrum contains two predominant features: 
(i) large-weighted structure consisting of the double sine curve, 
whose dispersions are approimately written as 
$\omega\sim(4t'\pm2t)\sin q$, and 
(ii) small-weighted continuum structure at low frequencies, 
which arises from excitations between different branches of 
the noninteracting bands.  A zero-energy excitation is caused by 
the creation of a particle-hole pair just at the Fermi level 
$\varepsilon_{\rm F}$, so that the gap closes at five momenta 
$q=0$, $k_{\rm F2}-k_{\rm F1}$ $(=\pi/4)$, $2k_{\rm F1}$, 
$2k_{\rm F1}$, and $k_{\rm F1}+k_{\rm F2}$.  

Now, using the DDMRG method, we attempt to reproduce the noninteracting 
spin excitation spectrum.  The result is shown in Fig.~\ref{fig1}.  
Since the noninteracting model poses a nontrivial problem to the DDMRG 
technique, it gives us a relevant accuracy test.  
When carrying out the DDMRG calculation, we have to take into account 
the required CPU time.  Usually, the DDMRG method takes much 
longer CPU time than the standard DMRG method because the excited 
states must be obtained and an asked quantity must be calculated 
(almost) individually for each frequency.  
Additionally, a required CPU time $\tau_{\rm CPU}$ increases rapidly 
with frequency $\omega$ and system size $L$ in the DDMRG calculation, 
which is estimated approximately as $\tau_{\rm CPU}\propto\omega^\alpha$ 
$(1<\alpha<2)$ and as $\tau_{\rm CPU}\propto L$ if we keep other 
conditions. 
Hence, it would be efficient to take a relatively small system for 
obtaining an overview of the spectrum and a larger system for 
studying the detailed structure at low frequencies.  

Let us then check our DDMRG result with the exact spectrum.  In the 
right panel of Fig.~\ref{fig2}, we show the spin excitation spectrum 
$S(q,\omega)$ at $U=0$ calculated with the DDMRG method in a chain 
with $L=24$ sites.  The structure of the double sine curve can be 
clearly seen.  However, it is hard to see dispersive structures in 
the continuum spectra at low frequenciese because only discrete peaks 
can be obtained due to the finite-size effect.  We need to take 
larger systems to resolve this problem since the resolution of 
spectrum can be improved in proportion to the system size.  We 
therefore choose to double the system size, $L=48$, and consider 
the low-energy excitations.  The result is shown in the left panel 
of Fig.~\ref{fig2}.  The resolution is obviously improved and we 
can now confirm the five momenta at which the zero-energy 
excitations occur.  
Moreover, we can see good agreement in the spin structure factor 
$S(q)$ $[=\sum_\omega S(q,\omega)]$ between the DDMRG and exact 
results, as shown in the inset of Fig.~\ref{fig2}.  
Thus, we are confident that our DDMRG method indeed enable us to 
study the detailed structures of the relatively complicated 
spectrum.  

For the information, we keep $m=400$ ($800$) density-matrix 
eigenstates to obtain the spectrum for $L=24$ ($48$) sites.  
Note that a larger $m$ value should be necessary to get the true 
ground state and excited state $s_{-q}^-|\Psi_0 \rangle$ 
(or $n_{-q}|\Psi_0 \rangle$) of the system.  
We therefore set $m=1200$ in the first $4-5$ DDMRG sweeps.  

\subsection*{Spin excitation spectrum}

\begin{figure}[t]
\begin{center}
\includegraphics[width= 7.0cm]{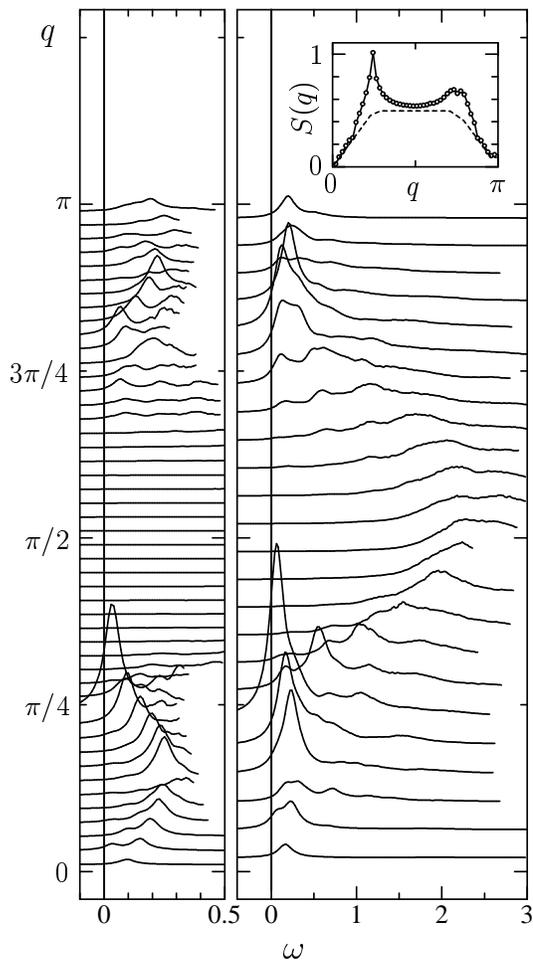}
\end{center}
\caption{
Spin excitation spectrum $S(q,\omega)$ at $t=0.25$, $t'=1$, 
and $U=10$ calculated by the DDMRG method.  We use the clusters 
$L=24$ with the broadening $\eta=0.1$ (right panel) and $L=48$ with 
$\eta=0.04$ (left panel).  Inset: Spin structure factor $S(q)$ 
obtained from $\omega$-integration of $S(q,\omega)$.}
\label{fig3}
\end{figure}

Next, let us see how the spin excitation spectrum is modified 
by the inclusion of the onsite Coulomb interaction $U$.  
In Fig.~\ref{fig3}, we show our DDMRG result for the spin 
excitation spectrum $S(q,\omega)$ at $t=0.25$, $t'=1$, 
and $U=10$, where we use the clusters with $L=24$ (right panel) 
and $L=48$ (left panel).  

Roughly speaking, the lower edge of the spectrum consists of 
three sine curves with four nodes at 
$q \sim 0$, $k_{\rm F2}-k_{\rm F1}$ $(=\pi/4)$, 
$\pi-(k_{\rm F2}-k_{\rm F1})$ $(=3\pi/4)$, and $\pi$.  
The excitation gap seems to close around these nodes.  This is 
consistent with previous theoretical studies,\cite{ohta,daul} 
which have suggested that no spin gap exists in the strong coupling 
regime.  On the other hand, the higher edge of the spectrum is 
approximately represented as a sine curve $\omega \sim \sin q$ 
as in the case of $U=0$, which comes from the creation of the 
particle-hole pairs within the same branch.  

Let us now take a closer look at the spectrm.  
As far as the spin degrees of freedom are concerned, the model 
Eq.~(\ref{hamiltonian}) for large $U$ may be mapped onto a two-chain 
$t$$-$$J$ model coupled with zigzag bonds.  For small $t/t'$, 
the antiferromagnetic interaction along the $t'$-chain, 
$J'$, must be much larger than that along the $t$-chain, $J$, 
if we assume that the exchange interaction comes from the 
second-order perturbation of the hopping integrals with a fixed $U$, 
i.e., $J'$ $(\sim t^{\prime 2})$ $\gg$ $J$ $(\sim t^2)$.  
Consequently, the features of the spectrum can be basically 
interpreted as those of the 1D quarter-filled $t$$-$$J$ 
model.\cite{bares,tohyama,saiga} 
The nodes at the lower edge of the DDMRG spectrum occur at the 
momenta $q=0$, $2k_{\rm F}^\ast$, $2\pi-2k_{\rm F}^\ast$, and $\pi$.  
Also, we can see considerable enhancement of spectral intensities 
around $q=\pi/4$ $(=k_{\rm F2}-k_{\rm F1})$ in comparison with the 
noninteracting spectrum.  This indicates that the onsite Coulomb 
interaction enhances the antiferromagnetic correlation with a 
period of $4$ times the lattice constant along the $t'$-chain, 
which can be easily expected from the fact that the $2k_{\rm F}^\ast$-SDW 
correlation is the most dominant for small $J$ in the 1D $t$$-$$J$ model 
at quarter filling.  This result may be compared with the 
$2k_{\rm F}$-SDW state observed experimentally in 
(TMTSF)$_2$X.\cite{ishiguro}

We then examine the effects of a small hopping integral $t$, 
which leads to the antiferromagnetic interaction $J$ along 
the $t$-chain as mentioned above.  From the viewpoint of the 
spin degrees of freedom, magnetic frustration must be brought 
because triangular lattices are formed of only the antiferromagnetic 
interactions.  Although $J$ is much smaller than $J'$, 
we can clearly see the influence in our DDMRG spectrum; 
there are two nodes around $q = 3\pi/4$, i.e., 
$q \sim 2k_{\rm F1}$ ($>3\pi/4$) and $2\pi-2k_{\rm F2}$ 
($<3\pi/4$).  These nodes are collected into a single node 
at $q = 3\pi/4$ when $t=0$.  
This split actually signifies a tendency to a formation of 
the $2k_{\rm F}$-SDW state along the $t$-chain as well as 
to a collapse of the $2k_{\rm F}$-SDW state along the $t'$-chain.  
With increasing $t/t'$, the node at $q=2k_{\rm F2}$ 
approaches $q=\pi/2$ and the adjacent spectral weight increases, 
whereas the node at $q=2k_{\rm F1}$ goes away from $q = 3\pi/4$ 
and the weight goes to zero.  
In other wards, the hopping term $t$ weakens the $2k_{\rm F}$-SDW 
oscillation along the $t'$-chain since the competing 
antiferromagnetic correlation along the $t$-chain is enhanced.  
In fact, the spectral weight around $q=\pi/4$ will certainly 
diminish as $t/t'$ increases.  

Another noticeable feature is that the spectral weight around 
$q=3\pi/4$ is obviously smaller than that around $q=\pi/4$, 
while the spectrum should be symmetrical about $q=\pi/2$ in 
the case of $t=0$.  This result implies the presence of 
ferromagetic spin fluctuations at low energies.  
For clearer understanding, we study the spin structure factor 
$S(q)$.  As shown in the inset of Fig.~\ref{fig3}, it is 
evident that $S(q)$ around $q=\pi/4$ is greater than that 
around $q=3\pi/4$; otherwise, $S(q)$ should be almost 
symmetrical about $q = \pi/2$.  This result indicates an 
enhancement of the ferromagnetic spin correlation between 
the neighbouring sites along the $t$-chain.  We may explain 
this as follows.  

A real-space behavior of the spin-spin correlation may be 
derived from the Fourier transform of $S(q)$, 
\begin{equation}
\left\langle S^z_i S^z_j \right\rangle=\frac{1}{L}\sum_qS(q)
e^{iq(r_i-r_j)},
\label{spinft}
\end{equation}
with $S^z_i=(n_{i\uparrow}-n_{i\downarrow})/2$.  We find 
that the spin correlation along the $t'$-chain is not 
affected so much by a small $t$ value but the decay length 
of the $2k_{\rm F}$-SDW oscillation is slightly shortened.  
On the contrary, the $t$ hopping term plays a prominent 
role in spin correlation between the $t'$-chains, 
which may be estimated as 
\begin{equation}
\left\langle S^z_i S^z_{i+R} \right\rangle\propto
\cos\big[(\pi/4)R\big]\times{\rm (decaying~term)}, 
\end{equation}
where $R$ is an odd number.  
For $R=1$, we obtain 
$\left\langle S^z_iS^z_{i+1}\right\rangle=0.00745$, 
which indicates the presence of ferromagnetic 
correlation between two electrons at the neighboring sites.  
This result is consistent with a scenario of spin-triplet 
superconductivity where the pairing of electrons occurs 
between the inter $t'$-chain nearest-neighbor sites, 
which is proposed in Ref.~\onlinecite{ohta}.  

\subsection*{Charge excitation spectrum}

\begin{figure}[t]
\begin{center}
\includegraphics[width=7.0cm]{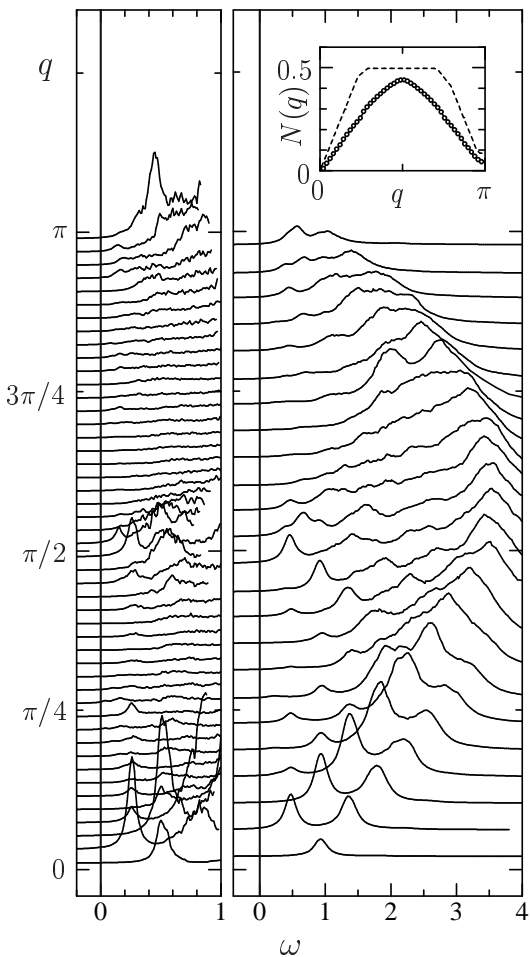}
\end{center}
\caption{Charge excitation spectrum $N(q,\omega)$ calculated by 
the DDMRG method for the same set of the parameter values as 
in Fig.~\ref{fig3}.  We use the clusters $L=24$ with broadening 
$\eta=0.1$ (right panel) and $L=48$ with $\eta=0.04$ (left panel).  
Inset: Charge structure factor $N(q)$ obtained from 
$\omega$-integration of $N(q,\omega)$.}
\label{fig4}
\end{figure}

Finally, we study the charge correlation function in the 
presence of the onsite Coulomb interaction $U$.  
In Fig.~\ref{fig4}, we show the charge excitation spectrum 
$N(q,\omega)$ at $t=0.25$, $t'=1$, and $U=10$, 
calculated by the DDMRG method for the chains with 
$L=24$ (right panel) and with $L=48$ (left panel).  
The outline of the spectrum, which is roughly expressed as 
$\omega \approx 4t' \sin q$, seems to be similar to 
that in the noninteracting spectrum.  This result reflects 
the fact that the overall dispersion is hardly affected by 
the onsite Coulomb interaction.  However, we notice that 
two distinct features emerge in the low-frequency range 
of the spectrum, which are discussed below.

One is the increase in the low-energy spectral weight around 
$q=\pi/2$ in the excitation spectrum.  The excitation gap seems 
to close there, so that the lower edge of the spectrum 
consistes of the two sine curves with three nodes at $q \sim 0$, 
$\pi/2$, and $\pi$.  We may thus have one gapless charge 
mode.  This result reflects a tendency toward 
the Peierls instability, i.e., the formation of the 
$4k_{\rm F}$-charge-density-wave (CDW) along the 
$t'$-chain.  We find that the spectral weight for the 
$2k_{\rm F}$-CDW correlation is instead much reduced.  
This feature would be more evident if we observe the 
momentum-dependent charge structure factor $N(q)$, which is 
shown in the inset of Fig.~\ref{fig4}.  We find that $N(q)$ 
takes the maximum value at $q=\pi/2$ and goes down along 
the practically straight lines to $q=0$ and $\pi$.  This 
dependence is consistent with results in the 1D quarter-filled 
Hubbard chain at large $U$.\cite{hirsch}; in fact, the result 
is almost equivalent to the result for noninteracting spinless 
fermions at half filling, as far as the charge 
degrees of freedom are concerned.  

The other is the apperance of large-weighted sharp peaks 
around $q=0$ at $\omega \approx 0$.  The point is as follows; 
the spectral weight of the peaks around $q=0$ is larger than 
that around $q=\pi$ in the low-frequency range, and they 
are also gathered at lower frequencies.  This implies that 
the electrons tend to come in the neighboring sites along 
the $t$-chain, which is associated with the pairing of two 
electrons between the inter $t'$-chain sites; 
accordingly, the pairs tend to be arranged alternately along 
the $t'$-chain.  Note that the two momenta $q=0$ and $\pi$ 
should be equivalent when $t=0$, so that the present result 
implies that the small hopping integral $t$ enhances the 
pairing fluctuations at low energies.  

Additionally, we can estimate the so-called Tomonaga-Luttinger 
liquid parameter $K_\rho$ from the derivative of $N(q)$ at $q=0$, 
\begin{equation}
K_\rho=\frac{\pi}{2}\frac{{\rm d}N(q)}{{\rm d}q}\Bigg|_{q=0}, 
\end{equation}
whereby we find the value $K_\rho \approx 0.637$.  
Because, in the presence of one gapless charge mode, the 
pairing correlation is dominant in comparison with 
the $4k_{\rm F}$-CDW correlation when 
$K_\rho>0.5$,\cite{nagaosa,schulz} 
our estimated value of $K_\rho$ is consistent with the 
occurence of spin-triplet superconductivity proposed in 
Ref.~\onlinecite{ohta}.

\section{SUMMARY}

We have calculated the spin and charge excitation spectra of 
the two-chain zigzag-bond Hubbard model at quarter filling in 
order to seek for consequences of the spin-triplet pairing 
induced by the ring-exchange mechanism.  
The model is topologically equivalent to the 1D Hubbard model 
with nearest-neighbor $t$ $(>0)$ and 
next-nearest-neighbor $t'$ $(>0)$ hopping integrals.  
We here have considered the case at $t\ll t'$.  
We have used the DDMRG technique to calculate the excitation 
spectra.  

We have first demonstrated the accuracy of the DDMRG method 
by reproducing the noninteracting exact spectrum.  
We have suggested that, for practical calculations, it is 
neccesary to adopt a relatively small system for obtaining 
an overview of the spectrum and a larger system for investigating 
detailed structures of the spectrum at low frequencies because 
the required CPU time increases rapidly with increasing the 
frequency and/or system size.  Thus, the DDMRG method enables 
us to study the details of relatively complicated structures 
of the spectra.

Then, we have investigated how the spectra are deformed when 
the strong onsite Coulomb interaction $U$ sets in.  
We find that the spin and charge excitation spectra are 
basically the same as those of the 1D Hubbard (and $t$-$J$) 
chain at quarter filling; i.e., the spectra come from the 
two nearly independent 1D chains where the $2k_{\rm F}$-SDW 
and $4k_{\rm F}$-CDW correlations along the $t'$-chains 
are enhanced. 
However, we find that the hopping integral $t$ plays a 
crucial role in the short-range spin and charge correlations; 
i.e., the ferromagnetic spin correlations between electrons 
on the neighboring sites is enhanced and the spin-triplet 
pairing correlations between the electrons is induced, 
of which the consequences are clearly seen in the calculated 
spin and charge excitation spectra at low energies.  
Our DDMRG calculations for the spin and charge excitation 
spectra have thus supported the ring-exchange mechanism for 
spin-triplet superconductivity where the pairing of electrons 
occurs between the two chains.  

\acknowledgments
We thank T.~Takimoto 
for useful discussions.  This work was supported in part by 
Grants-in-Aid for Scientific Research 
(Nos. 18028008, 18043006, and 18540338) 
from the Ministry of Education, Science, Sports, and 
Culture of Japan.  A part of computations was carried out 
at the Research Center for Computational Science, 
Okazaki Research Facilities, and the Institute for 
Solid State Physics, University of Tokyo.

\end{document}